\documentstyle[12pt]{article}
\begin{document}
.
\vspace{40mm}
\begin{center}
{\large \bf INTRODUCTION TO STRUCTURE FUNCTIONS \footnote {Introductory 
review talk presented at the XXXIst Rencontre de Moriond, "QCD and High Energy 
Hadronic  
Interactions", Les Arcs, France, 23 - 30 March 1996.}}\\
\vspace{5mm}
{\large \bf INTRODUCTION AUX FONCTIONS DE STRUCTURE}\\
\vspace{8mm}
J. Kwieci\'nski \\
Department of Theoretical Physics\\
H. Niewodnicza\'nski Institute of Nuclear Physics\\
Krak\'ow, Poland. \\
\end{center}
\vspace{60mm}
\begin{abstract}
The theory of deep inelastic scattering structure functions is reviewed 
with an emphasis put on the QCD expectations of their behaviour 
in the region of small values of the Bjorken parameter $x$.\\

  La th\'eorie des fonctions de structure de diffusion profondement 
        in\'elastique est pr\'esent\'ee  
	avec une attention particuli\`ere consacr\'ee aux pr\'evisions  
        de la Chromodynamique Quantique de leur comportement  
        dans la r\'egion des petites valeurs du param\`etre $x$ de Bjorken. 
\end{abstract}
\pagebreak 
The advent of the HERA $ep$ collider has opened up a possibility to test 
QCD in the new and hitherto unexplored regime of the small values of the 
Bjorken parameter $x$.  This parameter  
is, as usual, defined  as $x=Q^2/(2pq)$ where $p$ is the proton four momentum, 
$q$ 
the four momentum transfer between the leptons and $Q^2=-q^2$. 
  Perturbative QCD 
predicts that several new phenomena will occur when the parameter 
$x$ specifying 
the longitudinal momentum fraction of a hadron carried by a parton 
(i.e. by a quark or by a gluon) 
becomes very small \cite{GLR,BCKK} . The main expectation is
  that the gluon  densities 
should  strongly grow in this limit, eventually leading to the parton 
saturation effects \cite{GLR,BCKK,ADM1,JK1}. 
  The small $x$ behaviour of the structure functions 
is driven by the gluon through the $g \rightarrow q \bar q$ transition  
and the  increase of gluon distributions with decreasing $x$  
implies a similar increase of the deep inelastic lepton - proton  
 scattering structure function $F_2$ 
as the  Bjorken parameter $x$ decreases \cite{AKMS}.  
The recent experimental data are consistent with 
this perturbative QCD prediction that the structure function $F_2(x,Q^2)$ 
should strongly grow with the decreasing Bjorken parameter 
$x$ \cite{H1,ZEUS,PANITCH}.\\      

The purpose of this talk is to summarize the QCD expectations for the small $x$ 
behaviour of the deep inelastic scattering structure functions.  After briefly 
reviewing predictions of the Regge theory we shall discuss the Balitzkij, Lipatov, 
Fadin, Kuraev (BFKL) equation which sums the leading powers of 
$\alpha_s ln(1/x)$.  
We will also briefly describe the "conventional" formalism based on the 
leading (and next to leading) order QCD evolution equations and confront it 
with the BFKL equation.  Besides the  structure functions 
$F_2(x,Q^2)$ and $F_L(x,Q^2)$ we shall also consider the spin  
structure function $g_1(x,Q^2)$.  The novel feature in the latter case is  
 the appearence of the double logarithmic terms i.e. powers of 
$\alpha_sln^2(1/x)$ at each order of the perturbative expansion.\\

Small $x$ behaviour of structure functions  
for fixed $Q^2$ reflects the high energy behaviour of the 
virtual Compton scattering total cross-section with increasing   
total CM energy 
squared $W^2$ since $W^2=Q^2(1/x-1)$. The appropriate framework 
for the theoretical description 
of this behaviour is the Regge pole exchange picture  
\cite{PC} .\\

The high energy behaviour of the total 
hadronic and (real) photoproduction cross-sections can be economically 
described by two contributions: an (effective) pomeron with its 
intercept slightly above unity ($\sim 1.08$) and the leading meson 
Regge trajectories  
with  intercept $\alpha_R(0) \approx 0.5$ \cite{DOLA}.  
The reggeons can be 
identified as corresponding to $\rho, \omega$, $f$ or $A_2$ 
exchange(s) depending 
upon the quantum numbers involved. All these 
reggeons have approximately the same intercept.   
One refers to the pomeron obtained 
from the phenomenological analysis of  hadronic  total cross 
sections as the "soft" pomeron since the bulk of the processes building-up 
the cross sections are  low $p_t$ (soft) processes.\\

The Regge pole model gives the following parametrization of the deep 
inelastic scattering structure function $F_2(x,Q^2)$ at  
small $x$ 
\begin{equation}
F_2(x,Q^2)=\sum_i \tilde \beta_i(Q^2) x^{1-\alpha_i(0)}. 
\label{reggef}
\end{equation}
The relevant reggeons are 
those which can couple to two (virtual) photons.  The (singlet) part 
of the structure function $F_2$ is controlled at small $x$ by 
pomeron exchange, while the non-singlet part $F_2^{NS}=
F_2^p-F_2^n$ by the  $A_2$ reggeon.  
Neither pomeron nor $A_2$ reggeons  couple to the 
spin structure function $g_1(x,Q^2)$ which is described at small 
$x$ by the  exchange of reggeons corresponding to 
axial vector mesons \cite{IOFFE,EKARL} i.e. to  $A_1$ exchange for the non-singlet 
part $g_1^{NS} = g_1^{p}-g_1^{n}$ etc.
\begin{equation}
g_1^{NS}(x,Q^2)=\gamma(Q^2) x^{-\alpha_{A_1}(0)}. 
\label{gnsa1}
\end{equation}
  The  reggeons which correspond to axial vector mesons 
are expected to 
have very low intercept (i.e. $\alpha_{A_1} \le 0$ etc.).\\

At small $x$ the dominant role is played by the gluons 
and the  basic dynamical quantity is the   
unintegrated gluon distribution 
$f(x,Q_t^2)$ where $x$ denotes the momentum fraction 
of a parent hadron carried by a gluon and $Q_t$  its transverse 
momentum.  The unintegrated distribution $f(x,Q_t^2)$ 
is related in the following way to the more familiar scale dependent 
gluon distribution $g(x,Q^2)$: 
\begin{equation}
xg(x,Q^2)=\int^{Q^2} {dQ_t^2\over Q_t^2} f(x,Q_t^2). 
\label{intg}
\end{equation}
In the leading $ln(1/x)$  approximation the unintegrated 
distribution $f(x,Q_t^2)$ satisfies 
the BFKL equation \cite{BFKL,LIPATOV,CIAF} which has the following form: 
$$
f(x,Q_t^2)=f^0(x,Q_t^2)+
$$
\begin{equation}
\bar \alpha_s \int_x^1{dx^{\prime}\over 
x^{\prime}} \int {d^2 q\over \pi q^2}
\left[{Q_t^2 \over (\mbox{\boldmath $q$}+
\mbox{\boldmath $Q_t$})^2} 
f(x^{\prime},(\mbox{\boldmath $q$}+
\mbox{\boldmath $Q_t$})^2)-f(x^{\prime},Q_t^2)\Theta(Q_t^2-q^2)\right]
\label{bfkl}
\end{equation}
where 
\begin{equation}
\bar \alpha_s={3\alpha_s\over \pi}
\label{alphab}
\end{equation}
This equation sums the ladder diagrams with gluon exchange accompanied 
by virtual corrections which are responsible for the gluon reggeization. 
The first and the second 
terms  on the right hand side of  eq. (\ref{bfkl}) correspond 
to  real gluon emission with $q$ being the transverse 
momentum of the emitted gluon, and to the virtual corrections respectively. 
$f^0(x,Q_t^2)$ is a suitably defined inhomogeneous term.\\
 
For the fixed coupling case  eq. (\ref{bfkl}) can be solved 
analytically and the leading behaviour of its solution 
at small $x$ is given by the 
following expression:
\begin{equation} 
f(x,Q_t^2) \sim (Q_t^2)^{{1\over 2}} {x^{-\lambda_{BFKL}}\over 
\sqrt{ln({1\over x})}} exp\left(-{ln^2(Q_t^2/\bar Q^2)\over 2 \lambda^"
ln(1/x)} \right)
\label{bfkls}
\end{equation} 
with 
\begin{equation}
\lambda_{BFKL}=4 ln(2) \bar \alpha_s
\label{pombfkl}
\end{equation} 
\begin{equation} 
\lambda^"=\bar \alpha_s 28 \zeta(3) 
\label{diff}
\end{equation}
where the Riemann zeta function $\zeta(3) \approx 1.202$.  The 
parameter $\bar Q$ is of nonperturbative origin.\\

The quantity $1+ \lambda_{BFKL}$ is equal to the intercept of the so -  
called BFKL pomeron. Its potentially large magnitude ($\sim 1.5$) 
should be contrasted with the intercept $\alpha_{soft} \approx 1.08$ 
of the (effective) "soft" pomeron which has been determined 
from the phenomenological analysis of the high energy behaviour 
of hadronic and photoproduction total cross-sections \cite{DOLA}.\\   
  
The solution of the BFKL equation 
reflects its diffusion pattern which  
is the direct consequence of the absence of transverse momentum ordering 
along the gluon chain.   
The interrelation between the diffusion of transverse momenta towards 
both the infrared and ultraviolet regions {\bf and} the increase of gluon 
distributions 
with decreasing $x$ is a  characteristic property of QCD at low $x$.  
It has important consequences for the structure of the hadronic final 
state in deep inelastic scattering at small $x$ 
\cite{JK1,CONTRERAS,GRAF}.\\

In practice one introduces the running coupling $\bar \alpha_s(Q_t^2)$ 
in the BFKL equation (\ref{bfkl}). This requires the introduction of an 
infrared 
cut-off to prevent entering the infrared region where the 
coupling becomes large. The effective intercept $\lambda_{BFKL}$ 
found by numerically solving the equation depends   
on the magnitude of this cut-off \cite{KMS2}. The impact of the momentum 
cut-offs on the solution of the BFKL equation has also been discussed 
in refs. \cite{PVLC,MCDG}. In  impact parameter representation 
the BFKL equation offers an 
interesting  
interpretation in terms of colour dipoles \cite{DIPOLE}. The application 
of this formalism to the phenomenological analysis of deep inelastic scattering 
is presented in a talk given by Samuel Wallon \cite{WALLON}. It 
should also be emphasised that the complete calculation of the next-to-leading 
corrections to the BFKL equation has recently become presented in ref.
 \cite{NLX}.\\

The structure functions $F_{2,L}(x,Q^2)$ are driven  at small $x$ 
by the gluons 
 and are related in the following way to the unintegrated distribution $f$: 
\begin{equation}
F_{2,L}(x,Q^2)=\int_x^1{dx^{\prime}\over x^{\prime}}\int 
{dQ_t^2\over Q_t^2}F^{box}_{2,L}(
x^{\prime},Q_t^2,Q^2)f({x\over x^{\prime}},Q_t^2). 
\label{ktfac}
\end{equation}
The functions  $F^{box}_{2,L}(x^{\prime},Q_t^2,Q^2)$ may be regarded as  the 
structure 
functions of the off-shell gluons with  virtuality  
$Q_t^2$.  
They are described by the quark box (and crossed box) diagram contributions 
to the 
photon-gluon interaction.   
The small $x$ behaviour of the structure functions reflects the small 
$z$ ($z = x/x^{\prime}$) behaviour of the gluon distribution $f(z,Q_t^2)$.\\

Equation (\ref{ktfac}) is an example of the "$k_t$ factorization theorem" 
which relates measurable quantities (like DIS structure functions) to 
the convolution in both longitudinal as well as in transverse momenta of the 
universal gluon distribution $f(z,Q_t^2)$ with the cross-section 
(or structure function) describing the interaction of the "off-shell" gluon 
with the hard probe \cite{KTFAC,CIAFKT}.  
The $k_t$ factorization theorem is the basic tool for 
calculating the observable quantities in the small $x$ region in terms of the 
(unintegrated) gluon distribution $f$ which is the solution of the BFKL 
equation.\\ 

The leading - twist part of the $k_t$ factorization formula can be rewritten 
in a collinear factorization form.  The leading small $x$ effects are then 
automatically resummed in the  
 splitting functions and in the coefficient functions. The $k_t$ 
factorization theorem   can in fact be used as the tool for calculating 
these quantities.   Thus, for instance, the moment function $\bar 
P_{qg}(\omega, 
\alpha_s)$ of the splitting $P_{qg}(z, 
\alpha_s)$ function is represented in the following form 
(in the so called $Q_0^2$ regularization and DIS scheme \cite{CIAFKT}): 

\begin{equation}
\bar P_{qg}(\omega,\alpha_s)= 
{\gamma_{gg}^{2}({\bar \alpha_s\over \omega})\tilde F^{box}_{2}
\left(\omega=0,\gamma= \gamma_{gg}({\bar \alpha_s\over \omega})\right)
\over 2\sum_i e_i^2 }
\label{pqgf}
\end{equation} 
where  $\tilde F^{box}_{2}(
\omega,\gamma)$ is the Mellin transform of the moment function 
$\bar F^{box}_{2}(
\omega, Q_t^2,Q^2)$ i.e. 
\begin{equation}
\bar F^{box}_{2}(
\omega, Q_t^2,Q^2)={1\over 2 \pi i} \int_{1/2-i\infty}^{1/2+i\infty} 
d\gamma \tilde F^{box}_{2}(
\omega,\gamma)\left(Q^2\over Q_t^2 \right)^{\gamma}
\label{mbox}
\end{equation}
and the anomalous dimension $\gamma_{gg}({\bar \alpha_s\over \omega})$ 
has the following expansion \cite{JAR}; 
\begin{equation}
\gamma_{gg}({\bar \alpha_s\over \omega})=\sum_{n=1}^{\infty}c_n
\left({\bar \alpha_s\over \omega}\right)^n
\label{adexp}
\end{equation}
This expansion gives the following expansion of the splitting function 
$P_{gg} $
\begin{equation}
zP_{gg}(z,\alpha_s)=\sum_1^{\infty}c_n{[\alpha_s ln(1/z)]^{n-1}\over (n-1)!}
\label{pggzet}
\end{equation}
Representation (\ref{pqgf}) generates the following  
expansion of the splitting function $P_{qg}(z,\alpha_s)$ at small $z$:  
\begin{equation}
zP_{qg}(z,\alpha_s)={\alpha_s\over 2 \pi}zP^{(0)}(z) +
(\bar \alpha_s)^2 \sum_{n=1}^{\infty}b_n
{[\bar \alpha_s ln(1/z)]^{n-1}\over (n-1)!}
\label{zpqgf}
\end{equation}
  The first term on the right hand side of eq. (\ref{zpqgf})
vanishes at $z=0$.  It should be noted that 
 the  splitting function $P_{qg}$ 
is formally non-leading at small $z$ when compared with the splitting  
function $P_{gg}$ .   
For moderately small values of $z$ however,  
when the first few terms in the expansions (\ref{adexp}) and (\ref{zpqgf})
dominate, the BFKL effects can be much more important 
in $P_{qg}$  than in $P_{gg}$.  
This comes from the fact that in the expansion (\ref{zpqgf}) 
all coefficients $b_n$ are different from zero while in eq. (\ref{adexp}) 
we have $c_2=c_3=0$ \cite{JAR}.  
The small $x$ resummation effects within the conventional QCD evolution 
formalism have recently been discussed in refs. \cite{EKL,HBRW,BFORTE,FRT}. 
  One finds in general that at the moderately small 
values of $x$ which are relevant for 
the HERA measurements,   the small $x$ resummation effects in the 
splitting function $P_{qg}$ have a much stronger impact on $F_{2}$ than 
the small $x$ resummation in the splitting function $P_{gg}$. This  
reflects  the fact, which has already been mentioned 
above, that  in the expansion (\ref{zpqgf}) 
all coefficients $b_n$ are different from zero while in eq. (\ref{adexp}) 
we have $c_2=c_3=0$. It should also be remembered that the BFKL effects 
in the splitting function $P_{qg}(z,\alpha_s)$ can significantly affect 
extraction of the gluon distribution out of the experimental data on the 
slope of the structure function $F_2(x,Q^2)$ which is based on the following relation:
\begin{equation}
Q^2{\partial F_2(x,Q^2)\over \partial Q^2} \simeq 
2 \sum_i e_i^2 \int_x^1 dz P_{qg}(z,\alpha_s(Q^2)){x\over z}
g({x\over z},Q^2)
\label{slopef2}
\end{equation}
 
A more general treatment of the gluon ladder than that which follows 
from the BFKL formalism is  provided by 
the Catani, Ciafaloni, Fiorani, Marchesini (CCFM) equation based on
 angular ordering along the gluon chain 
\cite{CCFM,KMS1}.  
This equation embodies both the BFKL equation at small $x$ and the 
conventional Altarelli-Parisi evolution at large $x$.  
The unintegrated gluon distribution $f$ now acquires   
dependence upon an additional scale $Q$ 
which specifies the maximal angle of gluon 
emission.  
The CCFM equation has the following form : 
$$
f(x,Q_t^2,Q^2)=\hat f^0(x,Q_t^2,Q^2)+ $$
\begin{equation}
\bar \alpha_s
\int_x^1{dx^{\prime}\over 
x^{\prime}} \int {d^2 q\over \pi q^2} \Theta
(Q-qx/x^{\prime})\Delta_R({x\over x^{\prime}},Q_t^2,q^2)
{Q_t^2 \over (\mbox{\boldmath $q$}+
\mbox{\boldmath $Q_t$})^2} 
f(x^{\prime},(\mbox{\boldmath $q$}+
\mbox{\boldmath $Q_t$})^2,q^2))      
\label{ccfm}
\end{equation}
where the theta function $\Theta(Q-qx/x^{\prime})$ reflects the angular 
ordering constraint on the emitted gluon.  
The "non-Sudakov" form-factor $\Delta_R
(z,Q_t^2,q^2  )$ is now given by the following formula: 
\begin{equation}
\Delta_R(z,Q_t^2,q^2)=exp\left[-\bar \alpha_s\int_z^1 {dz^{\prime}
\over z^{\prime}} \int {dq^{\prime 2}
\over q^{\prime 2}}\Theta (q^{\prime 2}-(qz^{\prime})^2)
\Theta (Q_t^2-q^{\prime 2})\right]
\label{ns}
\end{equation}
Eq.(\ref{ccfm}) still contains only the singular term of the 
$g \rightarrow gg$ splitting function  
at small $z$. Its generalization which would  
include 
remaining parts of this vertex (as well as quarks) is possible.  
The numerical analysis of this equation was presented in ref. \cite{KMS1} 
. The CCFM equation which is the generalization of the BFKL equation generates 
the steep $x^{-\lambda}$ type of behaviour for the deep inelastic structure 
functions as the effect of the leading $ln(1/x)$ 
resummation \cite{CCFMF2}.  The slope $\lambda$ turns out to be sensitive 
on the (formally non-leading) additional constraint $q^2 < Q_t^2x^{\prime}/x$ 
in eq. (\ref{ccfm}) which 
follows from the requirement that the virtuality of the 
last gluon in the chain is dominated by $Q_t^2$ \cite{SAMUELSSON,KMSGLU}.\\

The HERA data can be described quite well using the BFKL and CCFM equations 
combined with the factorization formula (\ref{ktfac})  
\cite{AKMS,CCFMF2,KMSGLU}. One can however obtain satisfactory description of the HERA data staying 
within the scheme   
 based on the Altarelli-Parisi equations alone 
without the small $x$ resummation effects being included in the formalism 
\cite{MRS,GRV}.   
In the latter case the singular small $x$ behaviour of the gluon 
and sea quark distributions  
 has to be introduced in the parametrization of the starting 
distributions at the moderately large reference scale $Q^2=Q_0^2$   
 (i.e. $Q_0^2 \approx 4 GeV^2$ or so) \cite{MRS}.  One can also 
generate steep behaviour dynamically starting from  
non-singular "valence-like" parton distributions at some very low 
scale $Q_0^2=0.35GeV^2$ \cite{GRV}. In the latter case the gluon and sea 
quark 
distributions exhibit  "double logarithmic behaviour" \cite{DL} 
\begin{equation}
F_2(x,Q^2) \sim exp \left(2\sqrt{\xi(Q^2,Q_0^2)ln(1/x)}\right)
\label{dlog}
\end{equation}
where 
\begin{equation}
\xi(Q^2,Q_0^2)=\int_{Q_0^2}^{Q^2}{dq^2\over q^2}{3\alpha_s(q^2)\over \pi} . 
\label{evlength}
\end{equation} 
For very small values of the scale $Q_0^2$ the evolution length $\xi(Q^2,
Q_0^2)$  
can become large for moderate and large values of $Q^2$ and the "double 
logarithmic" behaviour (\ref{dlog}) is, within the limited region of $x$,  
similar to that corresponding to the power like increase of the type 
$x^{-\lambda}$, $\lambda \approx 0.3$.\\

The discussion presented above concerned the small $x$ 
behaviour of the singlet structure function which was driven by the gluon 
through the $g \rightarrow q \bar q$ transition. 
The gluons of course decouple from the non-singlet channel and the 
mechanism of generating the small $x$ behaviour in this case is different.\\

The novel feature of the non-singlet channel is the appearence of the 
{\bf double} logarithmic  terms i.e. powers of 
$\alpha_s ln^2(1/x)$ at each order of the perturbative 
expansion \cite{GORSHKOV,JK2,KL,EMR,BER}.  
These double logarithmic terms are generated by the 
ladder diagrams with  quark (antiquark) exchange along the chain.   
The ladder diagrams can acquire corrections from the "bremsstrahlung" 
contributions \cite{KL,BER}  
which do not vanish for the polarized structure function 
$g_1^{NS}(x,Q^2)$ \cite{BER}. They are  however relatively unimportant and 
are non-leading in the $1/N_c$ expansion. \\

In the approximation where the leading double logarithmic terms 
are generated by ladder diagrams with quark (antiquark) exchange along 
the chain  the 
unintegrated non-singlet spin dependent quark distribution $\Delta 
f_q^{NS}(x,k^2_t)$ 
($\Delta f_q^{NS}=\Delta u+\Delta \bar u - \Delta d -\bar \Delta d$) satisfies 
the following integral equation  :     
\begin{equation}
\Delta f_q^{NS}(x,Q_t^2)=\Delta f_{q0}^{NS}(x,Q_t^2)+ \tilde \alpha_s 
\int_x^1{dz\over z}\int_{Q_0^2}^{{Q_t^2\over z}} {dQ_t^{\prime 2}\over 
Q_t^{\prime 2}}\Delta f_q^{NS}({x\over z},Q_t^{\prime 2})
\label{dleq}
\end{equation}
where 
\begin{equation}
\tilde \alpha_s = {2 \over 3 \pi} \alpha_s 
\label{atil}
\end{equation} 
and $Q_0^2$ is the infrared cut-off parameter. 
The upper limit $Q_t^2/z$ in the integral equation (\ref{dleq}) follows 
from the 
requirement that the virtuality of the quark at the end of the chain 
is dominated by $Q_t^2$. A possible non-perturbative $A_1$ reggeon contribution 
has to be introduced in the driving term i.e. 
\begin{equation}
\Delta f_{q0}^{NS}(x,Q_t^2) \sim x^{-\alpha_{A_1}(0)}
\label{a2driv}
\end{equation}
at small $x$.\\
  
Equation (\ref{dleq}) implies the following equation 
for the moment function $\Delta \bar f_q^{NS}(\omega,Q_t^2)$ 
\begin{equation}
\Delta \bar f_q^{NS}(\omega,Q_t^2)=\Delta \bar f_{q0}^{NS}(\omega,Q_t^2)+ 
{\tilde \alpha_s \over \omega} \left[\int_{Q_0^2}^{Q_t^2} 
{dQ_t^{\prime 2}\over 
Q_t^{\prime 2}}\Delta \bar f_q^{NS}(\omega,Q_t^{\prime 2})+ 
\int_{Q_t^2}^{\infty} {dQ_t^{\prime 2}\over 
Q_t^{\prime 2}}\left({Q_t^2 \over Q_t^{\prime 2}}\right)^{\omega}
\Delta \bar f_q^{NS}(\omega,Q_t^{\prime 2})\right]
\label{dleqm}
\end{equation}
For fixed coupling $\tilde \alpha_s$  equation (\ref{dleqm})
 can be solved analytically.  
Assuming for simplicity that the inhomogeneous term is independent 
of $Q_t^2$ (i.e. that $\Delta \bar f_{q0}^{NS}(\omega,Q_t^2) = C(\omega)$ )
we get the following solution of  eq.(\ref{dleqm}): 
\begin{equation}
\Delta \bar f_q^{NS}(\omega,Q_t^2)=C(\omega)R(\tilde \alpha_s,  \omega) 
\left({Q_t^2\over Q_0^2}\right)^{ \gamma^{-}(\tilde \alpha_s,  \omega)}
\label{solm}
\end{equation}
where 
\begin{equation}
\gamma^{-}(\tilde \alpha_s, \omega) = {\omega - \sqrt{\omega^2 - 4 \tilde 
 \alpha_s}\over 2}
\label{anomd}
\end{equation}
and
\begin{equation}
R(\tilde \alpha_s,  \omega)= {\omega \gamma^{-}(\tilde \alpha_s, \omega)\over 
\tilde \alpha_s}. 
\label{r}
\end{equation}
Equation (\ref{anomd}) defines the anomalous dimension of the 
 moment of the non-singlet quark distribution in which 
 the double logarithmic $ln(1/x)$ terms i.e. the powers of ${\alpha_s \over 
\omega^2}$ have been resummed to all orders.  It can be seen from (\ref{anomd}) 
that this anomalous dimension has a (square root) branch point singularity 
at $\omega=
\bar \omega$ where 
\begin{equation}
\bar \omega= 2 \sqrt{\tilde \alpha_s}. 
\label{barom}
\end{equation} 
This singularity will of course be also present in the moment function $
\Delta \bar f_q^{NS}(\omega,Q_t^2)$ itself. It should be noted that in contrast to the 
BFKL singularity whose position above unity was proportional to $\alpha_s$,  
$\bar \omega$ is proportional to $\sqrt{\alpha_s}$ - this being the 
straightforward consequence of the fact that  equation (\ref{dleqm}) 
sums double logarithmic terms $({\alpha_s\over \omega^2})^n$. 
This singularity gives the following contribution to the 
non-singlet quark distribution $\Delta f_q^{NS}(x,Q_t^2)$ at small 
$x$:  
\begin{equation} 
\Delta f_q^{NS}(x,Q_t^2) \sim {x^{-\bar \omega}\over ln^{3/2}(1/x)}. 
\label{smxns}
\end{equation}

The introduction of the running coupling effects in  equation (\ref{dleqm})
turns the branch point singularity into the series of poles which accumulate 
at $\omega=0$ \cite{JK2}.  The numerical analysis of the corresponding 
integral equation,  
with the running coupling effects taken into account,  
gives an effective slope ,  
\begin{equation}
\lambda(x,Q_t^2)={dln \Delta f_q^{NS}(x,Q_t^2)\over d ln(1/x)}
\label{slope}
\end{equation}
with magnitude $\lambda(x,Q_t^2) \approx 0.2 - 0.3$ at small $x$ 
\cite{JK3}.   
The result of this estimate suggest that a reasonable extrapolation 
of the (non-singlet) polarized quark densities would be to assume an  
$x^{-\lambda}$ behaviour with $\lambda \approx 0.2 - 0.3$.   Similar 
 extrapolations of the spin-dependent quark 
distributions towards the small $x$ region have  
 been assumed in  several recent parametrizations of parton densities 
\cite{BS,GRVOG,GS,BV}.  
The perturbative QCD effects become significantly amplified for the 
singlet spin structure function due to  mixing with the gluons.  
The simple ladder equation may not however be  applicable 
for an accurate description of the double logarithmic terms in 
the polarized gluon distribution $\Delta G$ \cite{BERSING}.
The small $x$ behaviour of the spin dependent 
structure function $g_1$ has also been disscussed in refs. \cite{BASLO,RGRF}.
\\

To sum up we have briefly summarised the theoretical  QCD expectations for the 
structure functions at low $x$. We have limited ourselves 
to the  region of large $Q^2$ where perturbative QCD becomes applicable. 
Specific problems 
of the low $Q^2$, low $x$ region are discussed in ref. \cite{BBJK}. \\    

\medskip\medskip\medskip                
     
\section*{Acknowledgments}
I thank Jean Tr\^an Thanh V\^an for his kind invitation to the Moriond meeting and 
for very warm hospitality in Les Arcs. I thank Alan Martin and Peter Sutton 
for the very enjoyable research collaboration on problems presented 
in this review. 
This research has been supported in part by 
 the Polish State Committee for Scientific Research grant 2 P03B 231 08  and 
the EU under contracts n0. CHRX-CT92-0004/CT93-357.\\

\end{document}